\documentclass[conference,9pt]{IEEEtran}

\usepackage{cite}
\usepackage{amsmath,amssymb,amsfonts}
\usepackage{algorithmic}
\usepackage{gensymb}
\usepackage{graphicx}
\usepackage{subfigure}
\usepackage{textcomp}
\usepackage{xcolor}
\usepackage{amssymb}
\usepackage{graphicx}
\usepackage{amssymb}
\usepackage{amsmath}
\usepackage{color,rotating}
\usepackage{lmodern}
\usepackage{url}
\usepackage{cite}
\title{A Joint Design of MIMO-OFDM Dual-Function Radar Communication System Using Generalized Spatial Modulation}

\begin{document}

 \author{Zhaoyi Xu$^{1}$, Athina P. Petropulu$^{1}$ and Shunqiao Sun$^{2}$ \\ %\sthanks{Thanks to XYZ agency for funding.}}
	$^{1}$Department of Electrical and Computer Engineering, Rutgers University, Piscataway, NJ 08854 \\
	$^{2}$Department of Electrical and Computer Engineering, The University of Alabama, Tuscaloosa, AL 35487\\
	E-mail:  \texttt{\{zhaoyi.xu,athinap\}@rutgers.edu}, \texttt{shunqiao.sun@ua.edu}}

% \author{\IEEEauthorblockN{Zhaoyi Xu,Athina Petropulu}
% \IEEEauthorblockA{Department of Electrical and\\Computer Engineering\\
% Rutgers, The State University of New Jersey\\
% Piscataway, New Jersey 08854\\
% Email: zhaoyi.xu@rutgers.edu,athinap@soe.rutgers.edu}
% \and
% \IEEEauthorblockN{Shunqiao Sun}
% \IEEEauthorblockA{Department of Electrical and \\Computer Engineering\\
% The University of Alabama\\Tuscaloosa, Alabama 35487\\
% Email:  shunqiao.sun@ua.edu}}

\maketitle

\begin{abstract}

A novel dual-function radar communication (DFRC) system is proposed, that achieves high target resolution and high communication rate. It   consists of a  multiple-input multiple-output (MIMO) radar, where only a small number of antennas are active in each channel use.
The probing waveforms are  orthogonal  frequency division multiplexing (OFDM) type. The OFDM carriers are divided into two  groups, one that is used by the active antennas in a shared fashion, and another one, where each subcarrier  is assigned to an active antenna in an exclusive fashion (private subcarriers). Target estimation is carried out based on the received and transmitted symbols. 
The system communicates information via the transmitted OFDM data symbols and  the pattern of active  antennas in a generalized spatial modulation (GSM) fashion. A multi-antenna communication receiver can identify the indices of  active antennas via sparse signal recovery methods. The use of shared subcarriers enables high communication rate.
 The private subcarriers are used to synthesize a virtual array for  high angular resolution, and also for improved estimation on the active antenna indices. 
The OFDM waveforms allow the communication receiver to easily mitigate the effect of frequency selective fading, while the use of a sparse array at the transmitter reduces the hardware cost of the system.
The radar performance of the proposed DFRC system is evaluated via simulations, and bit error rate (BER) results for the communication system are provided.

% Our results suggest that with sharing subcarrier between activated antennas can provide a high bit rate while keeping a good resolution on target estimation. %numerical results of SNR and the probability of successful index recovery from
\end{abstract}


% \fi
\IEEEpeerreviewmaketitle

\section{Introduction}
Spectrum sharing between radar and communication systems aims at improving spectral efficiency. 
Dual-function radar-communication (DFRC) systems represent one form of spectrum sharing, by
providing radar and communication functionalities on  the same hardware platform \cite{dfrc,dfrcs}. Unlike  approaches that
consider spatially distributed radar and communication systems and require  coordination of the two system functions by some external controllers \cite{mc}, DFRC systems require less coordination. 
 A DFRC system periodically transmits probing waveforms that allow for  estimating target angle, range and velocity, and at the same time, convey information to a communication receiver. 
DFRC systems are  applicable in many scenarios, including  autonomous driving, where the radar functionality  can be used for  sensing and navigation \cite{adas},\cite{jrc,5gvv} and the communication functionality for vehicle to vehicle communication. 

Multiple-input multiple-output (MIMO) radar \cite{mimo} are good candidates for use in DFRC systems. 
They can form wide beams, thus allowing for the detection of multiple targets at the same time. %
% Compared with phased arrays, MIMO radar transmit different waveforms from their  transmit antennas. 
Further, when using  orthogonal waveforms, they  can synthesize  a virtual array that has a  larger aperture than that of a uniform linear array (ULA) with the same number of physical  elements. As a result, MIMO radar can achieve high angle resolution with a small number of antennas. 
 
The communication component of a
multi-antenna DFRC system can be realized by embedding information in the radar waveforms \cite{mimodfrc1,ofdm,ofdmdfrc}, or in the way  the waveforms are assigned to the transmit antennas \cite{phase,sm,smph}.
%
% \cite{ofdm}, showed that orthogonal frequency division multiplexing (OFDM) is a promising approach, and by operating solely on the modulation symbols  one can avoid the sidelobes that arise in OFDM radar processing.
% \textcolor{red}{why do you only comment on \cite{ofdm}? your presentation has to be balanced and classify the literature in different classes. What are the classes? }
% 
%
%
In \cite{ofdm,ofdmdfrc}, an orthogonal frequency division multiplexing (OFDM) DFRC system is proposed, where all OFDM  subcarriers are assigned to antennas in an exclusive fashion in order to maintain waveform orthogonality.  However, this use of subcarriers limits the communication rate of each antenna.
In \cite{sm},\cite{smph}, communication information is embedded in the transmit antenna activation pattern  by applying the generalized spatial modulation (GSM) idea of \cite{gsm}. 
  GSM has also been explored in MIMO radar \cite{phase}, by reconfiguring a sparse transmit array through antenna selection. However, 
 the amount of information that can be transmitted based on the active antenna pattern only is rather low.

In this paper, we propose a novel OFDM DFRC system, that achieves high target resolution and high communication rate. The proposed system   consists of a MIMO radar, where only a small number of antennas are active in each channel use.
 The OFDM carriers are divided into two  groups, one that is used by the active antennas in a shared fashion, and another one, where each subcarrier  is assigned to an active antenna in an exclusive fashion - those will be referred to as private subcarriers. Target estimation is carried out based on the received and transmitted symbols. 
The system communicates information via the transmitted OFDM data symbols, and the pattern of active transmit antennas, in a GSM fashion.
 A multi-antenna communication receiver identifies the indices of the active antennas via sparse signal recovery methods \cite{gsmcs1,gsmcs2}. 
pro

In comparison to \cite{ofdmdfrc} that also uses OFDM and multiple antennas, our proposed system uses subcarrier sharing, and thus achieves  higher communication rate.
The shared use of subcarriers results in coupling of angle, range and Doppler estimation. However, the synthesized  virtual array based on the private subcarriers allows for a high resolution  refinement of the initial angle estimate, which can subsequently yield a better estimate of target range and Doppler.
To the best of our knowledge, there are no other DFRC systems that have subcarriers in a shared fashion. 
% \textcolor{red}{As compared to \cite{smph}, which also ****DOES WHAT THIS HAVE IN COMMON WITH OUR METHOD???? is it OFDM based??*****, the virtual array on private subcarriers provides a higher resolution in angle estimation. }\textcolor{blue}{It uses GSM.} \textcolor{magenta}{Please email me \cite{smph}}

The remainder of this paper is organized as follows. Target information estimation is provided in Section II. Sections III describes the communication functionality of the proposed system. Section IV describes  how the radar and the communication functionalities work together. Section V demonstrates the performance of the proposed system via simulation results, and Section VI provides some concluding remarks.

\section{The proposed Radar system}

% \textcolor{red}{a full transmit ULA is not what you propose ***********maybe write this: ******}
%
% \textcolor{blue}{The proposed scheme uses a sparse transmit array, i.e., a ULA where only a subset of the transmit antennas are active in each channel use. 
% % For notational convenience and without loss of generality, in this section we present the radar system based on a full ULA.
% }
% \textcolor{brown}{The results for the  SLA  can be obtained by setting the symbols corresponding to inactive antenna  to  zero.  }
We consider a MIMO radar with a ULA transmit array with $N_t$ transmit elements, spaced apart by $d_t$, and a ULA receive array with $N_r$ receive elements, spaced apart by $d_r$.
In the transmit ULA, only $N_x$ antennas are active in each channel use. Let us denote by $\mathcal{N}$ the set of active antennas indices. We will assume that the $0$-th and the $(N_t-1)$-th elements always belong to $\mathcal{N}$, so that the aperture of the transmit array is fixed. 

The transmit waveforms are OFDM signals with $N_s$ subcarriers, with subcarrier spacing  $\Delta$. Each antenna applies an inverse discrete Fourier transform (IDFT) on the data symbols assigned to it, pre-appends a cyclic prefix (CP), converts the samples into an analog signal and transmits it with carrier frequency $f_c$. This signal will be referred to as an OFDM symbol.
 The length of the CP should be larger than the maximum roundtrip delay to the target, so that the inter-symbol interference and inter-channel interference can be eliminated in the following modulation symbol based radar processing.
% We will refer to the data symbols transmitted during one period as \textcolor{red}{\textit{OFDM symbol}}.

In the OFDM-MIMO radar of \cite{ofdmdfrc}, the subcarriers are distributed to the transmit antennas so that no two antennas transmit on the same subcarrier simultaneously. Here, we allow subcarrier sharing, which will enable a higher communication rate. In particular, we  divide the carriers into two groups, those who will be used in a shared fashion by the active antennas and the private subcarriers.

% Despite the fact that subcarrier sharing will disturb the orthogonality between transmitted waveforms, orthogonality still holds between private subcarrier which enables the synthesis of the virtual array.
% \textcolor{red}{comment on why they have not done this before. Is there a difficulty that you were the first to resolve?}

Let $d_{T_x}(n,i,\mu)$ denotes the data symbol transmitted by the $n$-th antenna ($n \in \mathcal{N}$) on the $i$-th subcarrier, during the $\mu$-th OFDM symbol. If subcarrier $i$ is a private subcarrier assigned to antenna $\ell$, then $d_{T_x}(n,i,\mu)\ne 0$ only if $n=\ell$. 
The baseband equivalent  of the corresponding transmitted waveform equals:
\begin{equation}
    \begin{split}
        x(n,t) = \sum_{\mu=0}^{N_p-1} \sum_{i=0}^{N_s-1} d_{T_x}(n,i,\mu) e^{j2\pi i\Delta t}rect(\frac{t-\mu T_p}{T_p}),
    \end{split}
    \label{xt}
\end{equation}  
 with $rect(t/T_p)$ denoting a rectangular pulse of duration $T_p$, where $T_p$ is the duration of OFDM symbol.

Suppose that there are $N_k$ point targets in the far field, each characterized by angle, range and Doppler frequency  $\theta_k, R_k,f_{d_k}$, respectively. It holds that $f_{d_k} = {2v_k f_c}/{c}$ with $c$ denoting the speed of light, and $v_k$  representing the velocity of the $k$-th target.
 The baseband equivalent of the signal reflected by the targets and received by the $m$-th antenna is
\begin{align}
        y(m,t)  &= \sum_{k=1}^{N_k}\sum_{n \in \mathcal{N}} x(n,t-\tau_k)e^{j2\pi f_{d_k}t}, 
        \label{receivedsig}
\end{align}
for $m = 0,...,N_r-1$, where
% where the Doppler effect arises due to the relative speed between the radar platform and the target and shifts the frequency of the received signal.
$\tau_k$ is the roundtrip delay of the $k$-th target, with  $\tau_k= 2R_k/c+(nd_t+md_r)sin\theta_k/\lambda_i$, and $\lambda_i = {c}/({f_c+i\Delta })$  the wavelength of the $i$-th subcarrier.

Each radar receive antenna samples in time, 
% with sampling period $1/(\Delta N_s)$, \textcolor{red}{you need to sample faster that 2 * max freq, i.e., faster that 2$\times \Delta N_s$}
% \textcolor{blue}{It is a complex signal, so the sampling frequency could be just the max freq}\textcolor{magenta}{the DFT is a sampled version of the DTFT with sampling interval $1/N_s$. But in order to have a correct DTFT you need to sample at Nyquist rate the continuous time signal }
discards the CP and applies an $N_s$-point discrete Fourier transform (DFT) on the samples to obtain  the symbols
\begin{align}
        d_{R_x}(m,i,\mu) =& \sum_{k=1}^{N_k}\sum_{n \in \mathcal{N}} d_{T_x}(n,i,\mu)
        e^{-j2\pi(md_r+nd_t)\sin\theta_k\frac{f_c+i\Delta }{c}} \nonumber\\
        &\times
        e^{-j2\pi i\Delta \frac{2R_k}{c}}e^{j2\pi \mu T_p f_{d_k}}.
        \label{prieq}
\end{align}
Eq.  \eqref{prieq} can be viewed  as
\begin{equation}
    d_{R_x}(m,i,\mu)=   \sum_{k=1}^{N_k} A(k,i,\mu) e^{j \omega(k,i)m}, \ m=0,...,N_r-1
    \label{fre_eq}
\end{equation}
where
\begin{align}
    A(k,i,\mu)=& \sum_{n \in \mathcal{N}} d_{T_x}(n,i,\mu)e^{-j2\pi nd_t\sin\theta_k\frac{f_c+i\Delta }{c}}
    \nonumber\\&\times e^{-j2\pi i\Delta \frac{2R_k}{c}}e^{j2\pi \mu T_p f_{d_k}}
    \label{ap1}
\end{align}
and
\begin{align}
    \omega(k,i) = -d_r\sin \theta_k \frac{f_c+i\Delta}{c} 
    \label{omega}
\end{align}
Assuming that $N_r>N_k$ and for a fixed $i$, $\{ d_{R_x}(m,i,\mu), m=0,...,N_r-1\}$ can be viewed as a sum of $N_k$ complex sinusoids with frequencies $\omega(k,i)$ and magnitudes $A(k,i,\mu)$. One can apply any of the existing methods  to find the frequencies and amplitudes of the sinusoids. For example, by applying an $N_r$-point DFT, we get peaks at frequencies $\omega(k,i)$. 
The resolution of the peaks will depend on the number of receive antennas, $N_r$.
Once  $\omega_k$ are estimated,  the target angles can be computed as
\begin{align}
    \theta_k = \arcsin \left({- \frac{\omega(k,i)c}{d_r(f_c+i\Delta)}}\right)
    \label{angle}
\end{align}

The amplitudes, $A(k,i,\mu)$, contain known data symbols and  target information, namely,  range and  Doppler. There can be multiple targets in the same angular bin. Suppose that there are $N_q$ targets at angle $\theta_{k}$. Then  the amplitude can be expressed as
\begin{align}
    A(k,i,\mu) &= \sum_{n \in \mathcal{N}} d_{T_x}(n,i,\mu)
        e^{-j2\pi nd_t\sin\theta_{k}\frac{f_c+i\Delta }{c}}\nonumber\\
        &\quad\times\sum_{q=1}^{N_q} e^{-j2\pi i\Delta \frac{2R_q}{c}} e^{j2\pi \mu T_p f_{d_q}}\nonumber\\
        &= A'(k,i,\mu)  \sum_{q=1}^{N_q} e^{-j2\pi i\Delta \frac{2R_q}{c}}e^{j2\pi \mu T_p f_{d_q}}
\end{align}
where $A'(k,i,\mu)=\sum_{n \in \mathcal{N}} d_{T_x}(n,i,\mu)e^{-j2\pi nd_t\sin\theta_k\frac{f_c+i\Delta }{c}}$. Via element-wise division  we get
%\textcolor{blue}{
\begin{align}
    d(k,i,\mu) \buildrel \triangle \over = \frac{A(k,i,\mu)}{A'(k,i,\mu)} = \sum_{q=1}^{N_q} e^{-j2\pi i\Delta \frac{2R_q}{c}}e^{j2\pi \mu T_p f_{d_q}}.
    \label{ele}
\end{align}
% \textcolor{blue}{***WRONG*****When the number of active antennas is odd and the indices of those antennas  are taken randomly in $[0,N_t-1]$, $A'(k,i,\mu)$ will be different than zero for a fixed $i$ and $\mu$, provided that the constellation diagram is usually centrosymmetric and the phase shifts are identical to different data symbols on the same subcarrier.*********}

% \textcolor{red}{regarding the division, you need to add comment in the paper - do not respond to me personally}

Eq. \eqref{ele} provides an expression that contains range and Doppler only, while the transmitted data have been eliminated. 
% \textcolor{blue}{which helps alleviating the high level sidelobes by eliminating the data dependency drawback in OFDM radars \cite{ofdm}.} 
 The range can then be estimated based on the peaks of an $N_s$-point  IDFT of $d(k,i,\mu)$, taken  along the $i$ dimension, i.e., 
%  \textcolor{red}{THIS IS PROBLEMATIC - we need to exclude the private subcarriers from the IDFT ******************** and this limits the resolution of range ****************** 
%  }
% \textcolor{blue}{No, we include the private ones in range estimation. Eq. 8-9 still hold on private subcarriers.} 
\begin{align}
    r(k,l,\mu) &= IDFT[d(k,i,\mu)] =
    \frac{1}{N_s}\sum_{i=0}^{N_s-1}d(k,i,\mu)e^{j\frac{2\pi}{N_s}il} \nonumber \\
    &=\sum_{q=1}^{N_q}\frac{e^{j2\pi \mu T_p f_{d_q}}}{N_s}
    \sum_{i=0}^{N_s-1}e^{-j2\pi i\Delta \frac{2R_q}{c}}
    e^{j\frac{2\pi}{N_s}il},
    \label{range}
\end{align}
for $l=0,...,N_s-1$.
 The peaks of $r(m,l,\mu)$ will appear  at  positions
 \begin{equation}
    l_q = \Big\lfloor\frac{2N_sR_q\Delta }{c}\Big\rfloor, 
    \label{rangeind}
 \end{equation}
where $\lfloor\cdot\rfloor$ denotes the floor function.
 %for $l=0,1,...,N_s-1$

 Similarly, by performing a discrete Fourier transform on \eqref{ele} along the dimension $\mu$,  we get peaks at 
%  \begin{equation}
%     l_k = \Big\lfloor\frac{2N_sR_k\Delta }{c}\Big\rfloor, 
%     \label{ranind}
%  \end{equation}
%  for $l=0,1,...,N_s-1$, where $\lfloor\cdot\rfloor$ denotes the round function.
%  Thus, at the indices from $l_1$ to $l_K$ in the $r(m,l,\mu)$, $K$ peaks will show up. Concurrently, the velocity of the targets can be extracted by deploy DFT on \eqref{r-d}:
%  \begin{align}
%  %   \begin{split}
%             v_{m}(i,p) &= DFT[d_m(i,\mu)] = \sum_{\mu=0}^{N_p-1}                     d_m(i,\mu)e^{-j\frac{2\pi}{N_p}p\mu} \nonumber\\
%             &=\sum_{k=1}^{N_k}e^{-j2\pi i\Delta \frac{2R_k}{c}}\sum_{n=0}^{N_t-1}e^{-j2\pi(nd_t+m d_r)\sin\theta_k\frac{f_c}{c}} \nonumber\\
%             &\times\sum_{\mu=0}^{N_p-1}e^{j2\pi \mu T_p f_{d_k}}
%             e^{-j\frac{2\pi}{N_s}p\mu}
%             \label{vel}
% %    \end{split}
% \end{align}
% for $p =0,1,...,N_p-1$.
%  It can also be seen that, the two exponential terms in \eqref{vel} cancel each other under the condition
 \begin{equation}
    p_q = \lfloor N_pT_pf_{d_q}\rfloor =\Big\lfloor\frac{2v_qf_cN_pT_p}{c}\Big\rfloor,  
    \label{veloind}
 \end{equation}
 for $p=0,1,...,N_p-1$. Based on the location of those peaks we can estimate the targets' velocities.
 
% \textcolor{green}{The performance of OFDM radar can be further improved by carefully designing the OFDM waveform. In \cite{sen2013designing},  a better detection performance and estimation accuracy can be achieved via designing the complex weights of OFDM waveforms where the multipath reflection exists. Targets can be exploited and resolved even outside the line of sight in multipath propagation scenarios which enables a more flexible detection of targets.  *************** THIS OPENS UP A NEW CAN OF WORMS - YOU HAVE NOT TOUCHED ON ANY OF THESE ISSUES IN THIS PAPER ***** DELETE - we need to address this in the journal paper}

% Ranges and Doppler frequencies of targets within a same angle bin can be individually but concurrently estimated based on the received signal at one of the received antennas, looking at the location of the peaks of the IDFT, taken along  dimension $i$, and the DFT, taken along dimension $\mu$. 
% Thanks to the fast Fourier transform (FFT), the detection of range-Doppler can be carried out fast and with low computation complexity.

% \subsection{On synthesizing a virtual array  \textcolor{red}{only one subsection, maybe remove the title??}}

\subsection{Angle estimation via virtual array synthesis} \label{sec_virtual}

The above presented angle estimation method, via \eqref{fre_eq}-\eqref{angle}, 
is based on an array of aperture $(N_r-1)d_r$, and the range and Doppler are coupled with angle. It turns out that once the angle is estimated and used to  obtain a range estimate, we can synthesize a virtual array to refine the angle estimate.
Here, we show how one can use the private subcarriers and the obtained range estimates to  synthesize a virtual array and achieve higher angle resolution.

The virtual array requires waveform orthogonality. To achieve that, let us assign a private subcarrier and antenna pairing $(i_n,n)$. 
There are $N_x$ private subcarriers at any time, where $N_x<N_s$.
Over the private subcarriers waveform orthogonality holds, and at the receiver, the contribution of each transmit antenna can be separated.  The symbol received by the $m$-th antenna on private subcarrier $i_n$ equals
\begin{align}
 %   \begin{split}
        d_{R_x}(m,i_n,\mu)=&\sum_{k=1}^{N_k}
        d_{T_x}(n,i_n,\mu)e^{-j2\pi(nd_t+md_r)\sin\theta_k\frac{f_c+i_n\Delta }{c}}
         \nonumber \\
        &\times e^{-j2\pi i_n\Delta \frac{2R_k}{c}}e^{j2\pi \mu T_p f_{d_k}}
        \label{virtual}
\end{align}
for $m = 0,1,...,N_r-1$ and $n\in \mathcal{N}$. 
% \textcolor{red}{maybe introduce the notation: $n\in \mathcal{N}$ where you define in the beginning of the paper $\mathcal{N}$ to be the set of active antenna indices, where the smallets is $0$ and the largest index is $N_t-1$ }
% where $\{n_1,n_2,...,n_{N_{ps}}\}$ are the indices of the antennas with private subcarrier and $n_id_t = iN_rd_r$. 
Provided that the spacing between subcarriers is much smaller as compared to $f_c$, we can approximate $f_c+i_n\Delta \approx f_c$. 
% SUN: for FDM MIMO radar, each TX antenna is corresponding to different $\lambda_i$ and because it is in the phase term for angle finding, I think this approximation is problematic. It is OK to approximate the phase difference among antennas or ignore them in range delay.
% \textcolor{blue}{Xu: Here the shift in phase we ignored is very small, and it can be tolerated by the angle resolution. Besides, the number of active antennas is small, so as the number of private subcarriers. Thus, this approximation is fine here.}
Then, after the element-wise division with the transmitted symbols, we get
\begin{align}
       d'(m,i_n,\mu)&{\buildrel \triangle \over =}  \frac{d_{R_x}(m,i_n,\mu)}{d_{T_x}(n,i_n,\mu)} \nonumber \\
        &=\sum_{k=1}^{N_k}
        e^{-j2\pi(n d_t+md_r)\frac{\sin\theta_k}{\lambda_0}}
        e^{-j2\pi i_n\Delta \frac{2R_k}{c}}e^{j2\pi \mu T_p f_{d_k}}
        \label{stvec}
\end{align}
for $m = 0,1,...,N_r-1$ and $n\in \mathcal{N}$.

Let $\alpha_{nk} = e^{-j2\pi nd_t\frac{\sin\theta_k}{\lambda_0}}e^{-j2\pi i_n\Delta \frac{2R_k}{c}}$ and $\beta_k =e^{j2\pi \mu T_p f_{d_k}}$. 
By stacking $d'(m,i_n,\mu)$ in vector ${\bf v}$,  \textcolor{black}{in an order that goes through all possible $m$'s for each $n\in \mathcal{N}$} we get
% \begin{equation}
%     {\bf v}=\sum_{k=1}^{N_k} \beta_k
%     \begin{bmatrix}
%     \alpha_{0k}\\\alpha_{0k}e^{-j2\pi d_r\frac{\sin\theta_k}{\lambda}} \\...
%     \\\alpha_{0k}e^{-j2\pi (N_r-1)d_r\frac{\sin\theta_k}{\lambda}}\\
%     \alpha_{\textcolor{red}{(N_1-1)}k}
%     \\...\\
%     \alpha_{(N_t-1)k} e^{-j2\pi (N_r-1)d_r\frac{\sin\theta_k}{\lambda}} 
%     \end{bmatrix}
%     \label{virvec}
% \end{equation}

\begin{align}
   {\bf v}= \sum_{k=1}^{N_k}\beta_k[{\bf D}(R_k)\odot{\bf a_t}(\theta_k)]\otimes {\bf a_r}(\theta_k)
   \label{kron}
\end{align}
where $\otimes$ is the Kronecker product, $\odot$ is the Hadamard product, ${\bf a_t}(\theta)=[1,e^{-j2\pi d_t\sin\theta/\lambda}, ..., e^{-j2\pi (N_t-1))d_t\sin\theta/\lambda}]^T$ and ${\bf a_r}(\theta)=[1,e^{-j2\pi d_r\sin\theta/\lambda}, ..., e^{-j2\pi (N_r-1))d_r\sin\theta/\lambda}]^T$ are the transmit and receive steering vector, respectively, and
\begin{align}
    {\bf D}(R) = {\bf I}_{\mathcal{N}}[
    e^{-j2\pi i_0\Delta \frac{2R}{c}},
    e^{-j2\pi i_1\Delta \frac{2R}{c}},...,
    e^{-j2\pi i_{N_t-1}\Delta \frac{2R}{c}}]^T
\end{align}
where ${\bf I}_{\mathcal{N}}$ is a diagonal matrix whose $n$-th diagonal element is $1$ if $n\in \mathcal{N}$, otherwise it is $0$.
% \textcolor{magenta}{On a second thought, by doing the above, we assume that we pair transmit antennas and private subcarriers, and then when we select the active antennas, the subcarriers paired with inactive antennas as wasted.
%
Eq. \eqref{kron} corresponds to a sparse ULA with aperture  $(N_t-1)d_t+(N_r-1)d_r$,  based on which, the targets parameters can be estimated via sparse signal recovery methods \cite{rossi2014}. 

Let ${R_1,R_2,...,R_{N_k}}$ be the already estimated target ranges. By discretizing the angle space on a grid of size $N_a$, i.e., $\{\tilde \theta(1), ..., \tilde \theta(N_a)\}$, Eq. \eqref{kron} can be expressed as
 \begin{eqnarray}
    {\bf v}&=&
      [{\bf v}_{11},{\bf v}_{12},...,{\bf v}_{N_a \times N_k}] \begin{bmatrix}
           \tilde\beta_{11} \\
           \vdots \\
           \tilde \beta_{N_a \times N_k}
            \end{bmatrix} \nonumber \\
            &=& [{\bf v}_{11},{\bf v}_{12},...,{\bf v}_{N_a \times N_k}]\tilde {\mathbf \beta}
 \label{virvec}
 \end{eqnarray}
 where $\tilde \beta_{ij}$ is non zero if there is a target at range $R_j$ and angle $\tilde \theta_i$ and \textcolor{black}{
 \begin{align}
     {\bf v}_{ij}=[{\bf D}(R_j)\odot{\bf a_t}(\theta_i)]\otimes {\bf a_r}(\theta_i)
 \end{align}
is the dictionary element for $i = 1,2,...,N_a$ and $j = 1,2,...,N_k$}. The sparse vector ${\bf \tilde \beta}$ can be estimated via $L_1$ norm minimization, and its support will provide target angle estimates.

\section{The Proposed Communication System}

In order to implement GSM, only  $N_x$ out of the $N_t$ ($N_x << N_t$) antennas will be active during a given transmission period. The indices of those antennas will change between transmission periods, and  will be used to encode information.
%
% is used to encode information bits, i.e., for each transmission round, only $N_x$ antennas will be active, while the indices of the active antennas will change between transmission rounds.
There are in total $C = {{N_t}\choose{N_x}} $
%\frac{N_t!}{N_x!(N_t-N_x)!}$ 
different active antenna selection possibilities. In each symbol period, those combinations will  result in 
\begin{equation}
    B = \lfloor\log_2(C)\rfloor \label{B}
\end{equation}
transmitted information bits.

The active antenna indices along with the transmitted data symbols can be estimated at the communication receiver as follows. Consider a communication receiver with $N_c$ antennas. 
%
% Due to the narrow bandwidth of each subcarrier frequency, the effect from a frequency selective fading channel between transmit antenna and receive antenna can be mitigated. Here we assume that the channel spread is smaller than the CP length.
% the channel between the $k$-th transmit antenna and the $j$-the receive antenna can be considered as a frequency selective fading 
% \textcolor{red}{BUT the whole point was to address frequency selective fading. This is why we used OFDM!} 
% The channel impulse response between the $k$-th transmit antenna and the $j$-th receive antenna is denoted as  $h_{j,k}(n), n=0,...,L-1$.  Let us assume that the CP is larger than $L$. 
% 
% 
% 
% \textcolor{red}{the comm receiver needs to know the length of CP in order to demodulate. It also needs to be synchronized with he radar transmitter}.
The received symbol matrix corresponding to the $\mu$-th OFDM symbol equals
\begin{equation}
{\bf Y} = 
    \begin{bmatrix}
    d_{C_x}(0,0,\mu)&...&d_{C_x}(0,N_s-1,\mu)\\
    d_{C_x}(1,0,\mu)&...&d_{C_x}(1,N_s-1,\mu)\\
    ...&...&...\\
    d_{C_x}(N_r-1,0,\mu)&...&d_{C_x}(N_r-1,N_s-1,\mu)\\
    \end{bmatrix}
\end{equation}
% \begin{equation}
%     d_{C_x}(m,i) = {\bf H}_{m,i}{\bf A_i} = \sum_{n=0}^{N_t-1}H(m,n,i)d_{T_x}(n,i)
% \end{equation}
% where ${\bf H}_{m,i}$ and ${\bf A_i}$ denotes the $m$-th column of the $i$-th layer of the channel matrix ${\bf H} \in\mathbb{C}^{N_c\times N_t\times N_s}$ and ${\bf A}_{i}=  [d_{T_x}(1,i),d_{T_x}(2,i),...,d_{T_x}(N_t,i) ]^T$ is the $i$-th column of symbol matrix ${\bf A}\in\mathbb{C}^{N_t\times N_s}$.
where $d_{C_x}(m,i,\mu)$ refers to the complex symbol received by the $m$-th communication receive antenna on the $i$-th subcarrier. As a result of subcarrier sharing and the  narrow bandwidth of the OFDM subcarriers, the $i$-th column of $\bf Y$ can be expressed as
\begin{equation}
\begin{split}
        {{\bf Y}_i}
        ={{\bf H}}_i{{\bf A}_i} + {\bf N}, \ i = 0,...,N_s-1
\end{split}
\end{equation}
where ${\bf H}_i \in\mathbb{C}^{N_c\times N_t}$
 is the frequency response of the channel between the transmit and receive antennas along the $i$-th carrier;
% ${\bf H}_{j,k}\in \mathbb{C}^{1\times1\times N_s}$ denotes the channel frequency response between the $k$-th transmit antenna and the $j$-the receive antenna and is determined from a $N_s$-point DFT of $h_{j,k}(n)$.
% \textcolor{red}{define the elements of this matrix in terms of the  $N_s$-point DFT of $h_{j,k}(n)$}
%
% containing the channel between the $j$-th receiver and the $k$-th transmitter at the $i$-th subcarrier frequency $h_{j,k}(i)$, 
% If ${\bf A}\in\mathbb{C}^{N_t\times N_s}$ denotes the symbol matrix containing the data symbols contained in one OFDM symbol, then 
${\bf A}_{i}=  {\mathbf{I}_{\mathcal{N}}}[d_{T_x}(1,i,\mu),d_{T_x}(2,i,\mu),...,d_{T_x}(N_t,i,\mu) ]^T$ 
containing the  data symbols transmitted on the $i$-th subcarrier; and   ${\bf N} \in\mathbb{C}^{N_c}$ is  additive white Gaussian noise.

\subsection{Information recovery via sparse signal recovery methods}\label{SSR}
When only a small fraction of the radar transmit antennas is active at a time, ${\bf A}_i$ will be sparse.
For a given $\mu$, all ${\bf A}_i$'s for $i=0,...,N_s-1$ have the same sparsity pattern.
% \textcolor{blue}{since we enable the sharing of subcarriers between active antennas.}
%provided that the active antennas are the same. 
Then, under certain conditions, ${\bf A}_i$ can be recovered by solving a sparse signal recovery problem \cite{gsmcs1},\cite{gsmcs2}. 
% Such recovery requires the number of receive antennas \textcolor{red}{to at least two times the number of active antennas \cite{cs} **** are you sure this is what \cite{cs} says? I have never seen such condition.}\textcolor{blue}{Under this condition, the sparse problem will have an unique sparse solution. I should add this.}\textcolor{red}{I doubt there is such condition. Please check with Shunqiao}\textcolor{blue}{This requires another precondition. I will remove this, it is not suitable here.}
%
By applying the same process to every subcarrier and every OFDM symbol,   all transmitted symbols can be recovered. The support of the recovered ${\bf A}_i$ provides the active antenna indices.
By decoding those indices  the transmitted bits in one period can be increased by $B$.

Compared with an OFDM communication system with the same modulation scheme but without subcarrier sharing, the proposed scheme increases the number of information bits transmitted in  one period by a factor of $N_x$ at maximum.
% Generalized spatial modulation(GSM) is a technique that we can encode the information onto both the constellation diagram and the position of activated antenna. Based on the position bits and encoding mechanism, we choose the antennas to be activated. At the receiver side, the receiver can find the position of the activated antennas as well as the information symbols. 
% By exploiting the sparse structure of the transmit antenna array,  we can recover the symbols jointly with the indices of the antennas that transmitted them. Thus, by decoding the position bits from the dictionary, the transmitted bits in one period can be increased by $B$.
% Moreover, the position indices comes freely from the recovered symbol vector $x$.

% Beyond normal generalized spatial modulation, the symbols are recovered from each subcarrier, which enables encoding information at each subcarrier frequency. In other words, we can do generalized spatial modulation at each subcarrier frequency on the activated antennas. The whole symbol matrix will have several columns have non-zero elements and along these columns, only some rows of them are selected to be non-zero. Take the computation demand of solving convex question into account, we can have a relative low hardware cost while compensating the reduced bit rate. 
\subsection{Information recovery by exploiting the private subcarriers}\label{alternative}

Here we provide an alternative way to estimate the transmitted symbols, by exploiting the private subcarriers.

% The active antenna indices can be recovered based on the private subcarriers as follows.
% \textcolor{red}{how does the comm receive know which are the private subcarriers?  explain}
% \textcolor{blue}{Xu: by the following step to label the active antenna indices.}
% An alternative way to compute the active antennas indices, by focusing on the private subcarriers, is the following.
% \textcolor{red}{SSR is significantly more complex. You need to explain why you still propose it and not stick with the method below}\textcolor{blue}{We need SSR on private subcarriers.}
% We propose to assign a private subcarrier to each active antenna for the recovery of active antenna indices.
% The communication receiver will apply the same sparse sensing method to all the subcarriers and all OFDM symbols. 

The communication receiver does not know which subcarriers are  private. However, if after applying sparse signal recovery on a certain subcarrier
the recovered sparse vector contains only one nonzero element, then the receiver may conclude that that subcarrier was private and  the non-zero symbol location  corresponds to the index of the active antenna matched to that subcarrier.
In that way, the receiver can identify all private subcarriers and active antenna indices. Subsequently, the receiver can  estimate the transmitted symbols on the  shared subcarriers via  least-squares estimation. 
\textcolor{black}{As it will be shown in the simulations section, this approach is more robust than estimating the symbols via the method of Section~\ref{SSR}.}

% \textcolor{red}{I would DELETED THE FOLLOWING - I do not understand what you say here and I do not think that any speial attention is needed ***** Thus, enjoy this benefit, the private subcarriers should be placed adjacently at the head of shared subcarriers such that the communication receiver can find it in the first place to determine the active antenna indices. }

\section{Dual-Function System}\label{dual}
In this section we  discuss how the radar and communication components of the system are implemented.

\underline{Radar transmitter:}
The  bit stream is divided into multiple sections, each section containing the symbols to be assigned to each antenna, i.e., the symbols comprising the OFDM symbol to be transmitted by the antenna. Each section is preceded by $B$    bits, indicating the indices of antennas to be active. The indices of active antennas change between channel uses.

\underline{Radar receiver:}
The angles are estimated by first performing an $N_r$-point  DFT on \eqref{fre_eq} along the $m$ dimension. The location of the peaks are the frequencies  of \eqref{omega}, which then lead to the target angles via  \eqref{angle}.
Subsequently, the target ranges are  estimated based on \eqref{rangeind}, and  the  velocities  based on \eqref{veloind}. 
\textcolor{black}{
To maintain full range resolution, in each OFDM symbol, the full bandwidth should be used to carry symbols.}
% provided that in \eqref{rangeind}, the resolution is decided by the total bandwidth. 
Similarly, to maximize the Doppler resolution, at least one subcarrier should be modulated with data symbols in all OFDM symbols, since the Doppler resolution  is determined by the total time of observation on the subcarrier.

\underline{Using the  virtual array:} The angle estimates can be refined  along the lines of Section~\ref{sec_virtual}, and 
 used to  improve the estimation  of  range and Doppler. The latter can be done evaluating \eqref{ele}  with the refined frequency estimates and then repeating the   range-Doppler estimation.

For the sake of achieving a virtual array with maximum aperture, the first and last active antennas need to be fixed, which slightly reduces the number of antenna activation patterns.

\underline{About the private carriers:}
The use of private subcarriers comes at the cost of limiting the spatial encoding and losing $N_x(N_x-1)$ data symbols.
In order to reduce the loss, and if the target is not changing fast over $M$ OFDM symbols ($M<N_p$), we can use private subcarriers only once every  $M$ OFDM symbols.
In most scenarios this is a reasonable assumption. For example, for an OFDM system with subcarrier bandwidth $100$kHz, the OFDM symbol duration is $10\mu$s.
For an array with $1\degree$ angle resolution, for a target at $50$m to move out of the angle bin it would require speed of $\frac{8.73}{M}\times 10^4$m/s; this means that the target will stay in the same angle bin for several OFDM symbols.

\underline{Information required:}
 In the proposed DFRC system, the communication receiver  needs to know the channel matrix $\bf H$, the number of subcarriers and the length of CP. %the number of transmit antennas, $N_x$, 
 Synchronization at symbol level is also assumed.

\section{Simulation Results}

 In this section, we demonstrate via simulations the radar target detection and communication performance of the proposed DFRC system. 
 
 The channels are simulated to be frequency selective and the corresponding impulse  responses are complex with zero-mean jointly Gaussian real and imaginary parts. 
The  system parameters are shown in Table \ref{table1}.
The antennas transmit $16$-QAM
signals in an OFDM fashion.
 Several point targets in the far field of the array
 are considered, each characterized by (angle, range, velocity), with values as shown in Table~\ref{radar}.

Based on Table \ref{radar},   two of the targets have the same relative velocity of $5$m/s and  are closely placed, i.e., have  polar coordinates  ($19\degree$,$50$m) and ($22\degree$,$50$m). 
 In order to construct a virtual array, the first $N_x$ subcarriers are set as private.

% \textcolor{red}{Explain how you simulated the channels.}

 %We simulated an OFDM radar system and a MIMO communication systems based on the parameters shown in Table \ref{table1}.
 
%  \textcolor{red}{why so many receive antennas at the radar?
%  How many antennas for the comm system? The number of antennas should be varied and see the effect.}
 
% Our simulations demonstrate the ability of the proposed OFDM based DFRC to identify targets that fall within the same range-Doppler bin, and targets that appear at the same direction of arrival (DOA). 
 
\begin{table}[!h]
\caption{Radar Parameters}
\label{table1}
\centering
\resizebox{75mm}{18mm}{
\begin{tabular}{ |c||c|c|  }
 \hline
 Parameter & Symbol & Value\\
 \hline
 Center frequency   & $f_c$    &24GHz\\
 Subcarrier spacing &   $\Delta $  & 100kHz\\
 Cyclic prefix length & $T_c$ & 2.5$\mu$s\\
 Duration of  OFDM symbol & $T_p$ & 12.5$\mu$s\\
%  Duration of modulated OFDM symbol & $T$ & 12.5$\mu$s\\
 Number of subcarriers & $N_s$ & 1024\\
 Number of OFDM symbols & $N_p$ & 256\\
 Total number of transmit antennas & $N_t$ & 32\\
 Number of activated antennas & $N_{x}$ & 5\\
 Number of radar receive antennas & $N_r$ & 50\\
 Number of communication receive antennas & $N_c$ & 16\\
 Receive antenna spacing distance & $d_r$ & 0.5$\lambda$\\
 Transmit antenna spacing distance & $d_t$ & 1$\lambda$ \\
 \hline
\end{tabular}}
\end{table}

%
 
%  at different locations with different relative velocities
% are assumed. 
% \textcolor{red}{The target angles are assumed to be identical for all transmit and receive antennas, *****NOT SURE WHAT THIS MEANS ????}
% and the targets stay in the same angle and range-Doppler bin during the measurement.

% \begin{figure}%[t]
% \centering
% \subfigure[]{\includegraphics[width=3.0in]{S-rd-R.pdf}} \\%
% \subfigure[]{\includegraphics[width=3.0in]{S-rd-I.pdf}}%
% \caption{Target estimation of four point targets where two of them are within the same range-Doppler bin. }
% \label{radar}
% \end{figure}

\begin{table}[!h]
\caption{Radar Parameters}
\centering
% \resizebox{75mm}{18mm}
{
\begin{tabular}{ |c|c| }
 \hline
Target parameters & Estimated  parameters \\
\hline
($19\degree$,$50$m,$5$m/s) &  ($19\degree$,$49.80$m,$5.86$m/s)\\
($7\degree$,$45$m,$10$m/s) &  ($7\degree$,$45.41$m,$9.77$m/s)\\
($19\degree$,$80$m,$7$m/s )&  ($19\degree$,$80.57$m,$7.81$m/s)\\
($22\degree$,$50$m,$5$m/s) & ($22\degree$,$49.80$m,$5.86$m/s)\\
\hline
\end{tabular}}
\label{radar}
\end{table}

The radar first estimates the  target angles via the low-resolution method (eqns. \eqref{fre_eq}-\eqref{angle}), and  then  estimates the target ranges corresponding to each angle, based on Eq. \eqref{rangeind}. {The obtained  estimates are ($6.89\degree$,$45.41$m,$9.77$m/s), ($18.66\degree$,$49.80$m,$5.86$m/s) and ($18.66\degree$,$80.57$m,$7.81$m/s), where one can see that  one target has not been resolved.} Based on the estimated target ranges, a high-resolution angle estimate is obtained based on the virtual array. The obtained parameters 
 are given in Table~\ref{radar}, where one can see that the closely spaced targets have been resolved.

\begin{figure}%[!htbp]
\centering
\includegraphics[width=2.9in]{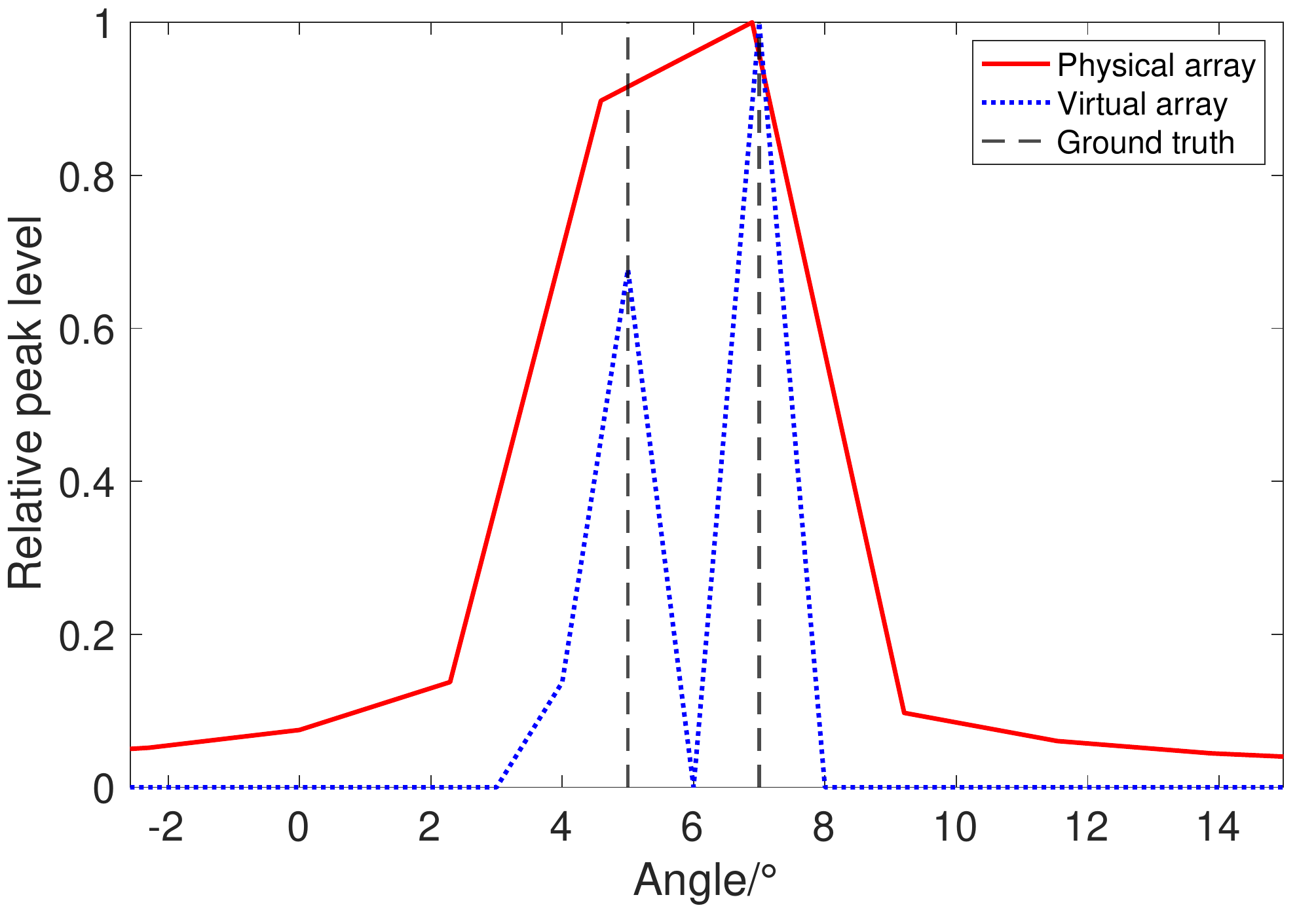}
\caption{Angle estimation can be improved with the virtual array. The targets are at 5$\degree$ and 7$\degree$.}% the apertures of receive array and virtual array are 4.9$\lambda$ and 34.3$\lambda$ respectively. The targets are at 19$\degree$ and 22$\degree$ respectively.}
\label{vir}
\end{figure}

% \textcolor{red}{GLOBAL: COMMENT: all numbers should appear in math more (e.g., $3$)}

% \textcolor{red}{make fonts on figures larger}
% \textcolor{red}{SUN: For all figures, you can save the figures to pdf files in Matlab, and then remove their white edge space in acrobat; better to use pdf files rather than jpg.}

% The angle resolution depends on the array aperture.  Even when the aperture of the physical array is not large enough to satisfy the angular resolution requirement, the virtual array has larger aperture and can achieve a higher resolution.
% Even if the total number of receive antennas is not large enough to satisfy the angular resolution requirement, the virtual array has larger aperture and achieves higher resolution.
Fig. \ref{vir} demonstrates the improvement in angular resolution enabled by the virtual array. 
In this case there are two targets, at $5\degree$ and $7\degree$ angles and both at the same range. 
%  the previous aperture of the receive array is $4.9\lambda$ and the aperture of virtual array is $24.5\lambda$. 
   {The red line shows the magnitude DFT of 
   $\{ d_{R_x}(m,i,\mu), m=0,...,N_r-1\}$,  with the location of the peaks indicating the angle estimates. The aperture of the receive array, i.e.,  $(N_r-1)d_r=24.5\lambda$,  does not allow for the estimation of  closely placed targets and thus only one peak shows up. The blue line shows the magnitude of $\tilde \beta$ of \eqref{virvec}, with the peaks indicating 
   the high resolution angle estimates. The virtual array  is a sparse version of an array of aperture $(N_t-1)d_t +  (N_r-1)d_r=55.5\lambda$ },  thus, due to its higher resolution   the targets can be resolved.  For the virtual array based estimate, a grid of size $N_a = 181$ was used. In this simulation, no noise was added.

\begin{figure}
\centering
\includegraphics[width=3in]{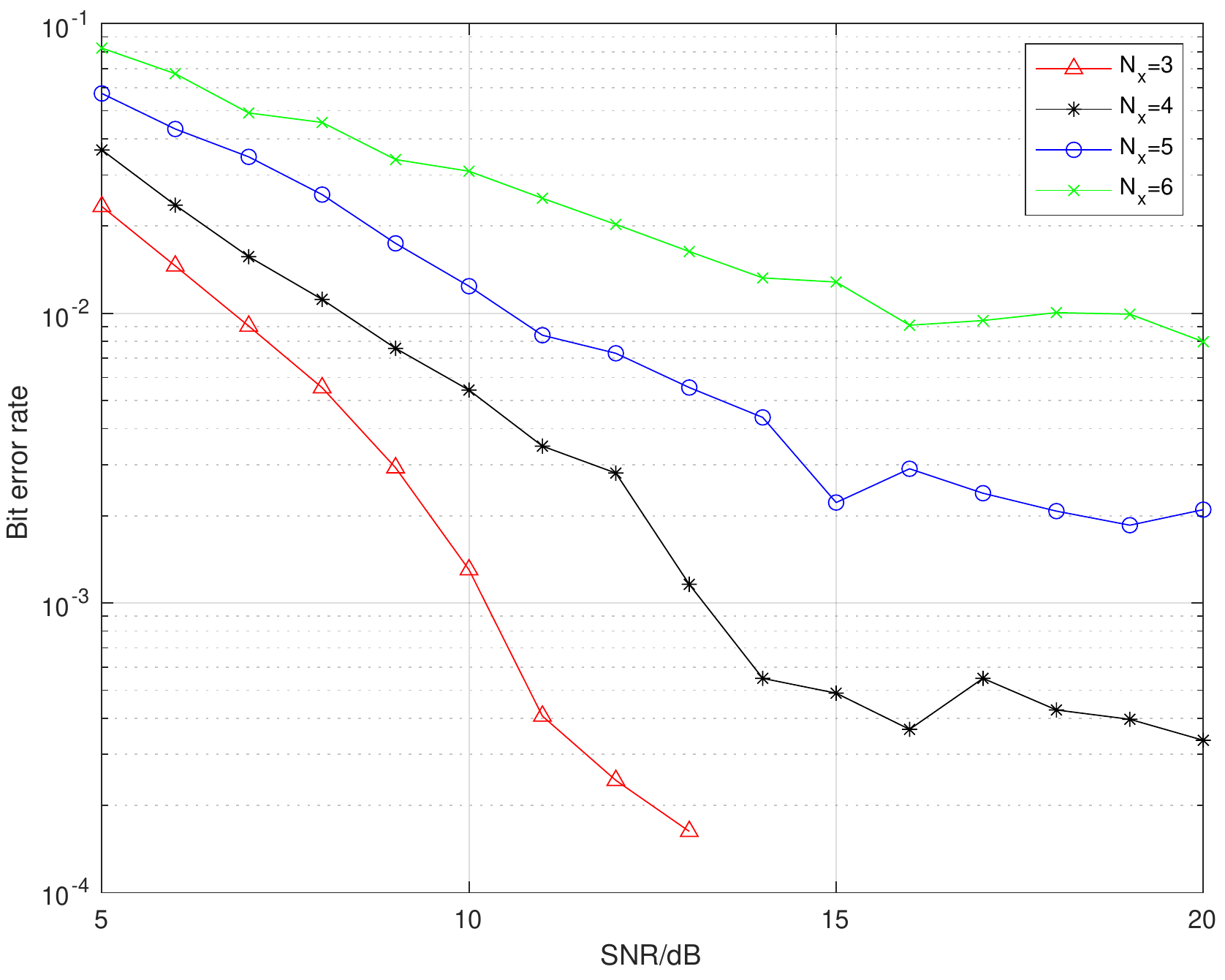}
\caption{BER versus SNR based on data symbols obtained via sparse signal recovery.}

% \textcolor{blue}{in data symbols on shared subcarriers using sparse signal recovery.}\textcolor{red}{without exploiting information in the antenna indices **** how would exploiting that info would help BER?}\textcolor{blue}{Xu: reduce the problem to least-square}
% }
% \textcolor{red}{the graph says probability}}
\label{ber}
\end{figure}

\begin{figure}
\centering
\includegraphics[width=3in]{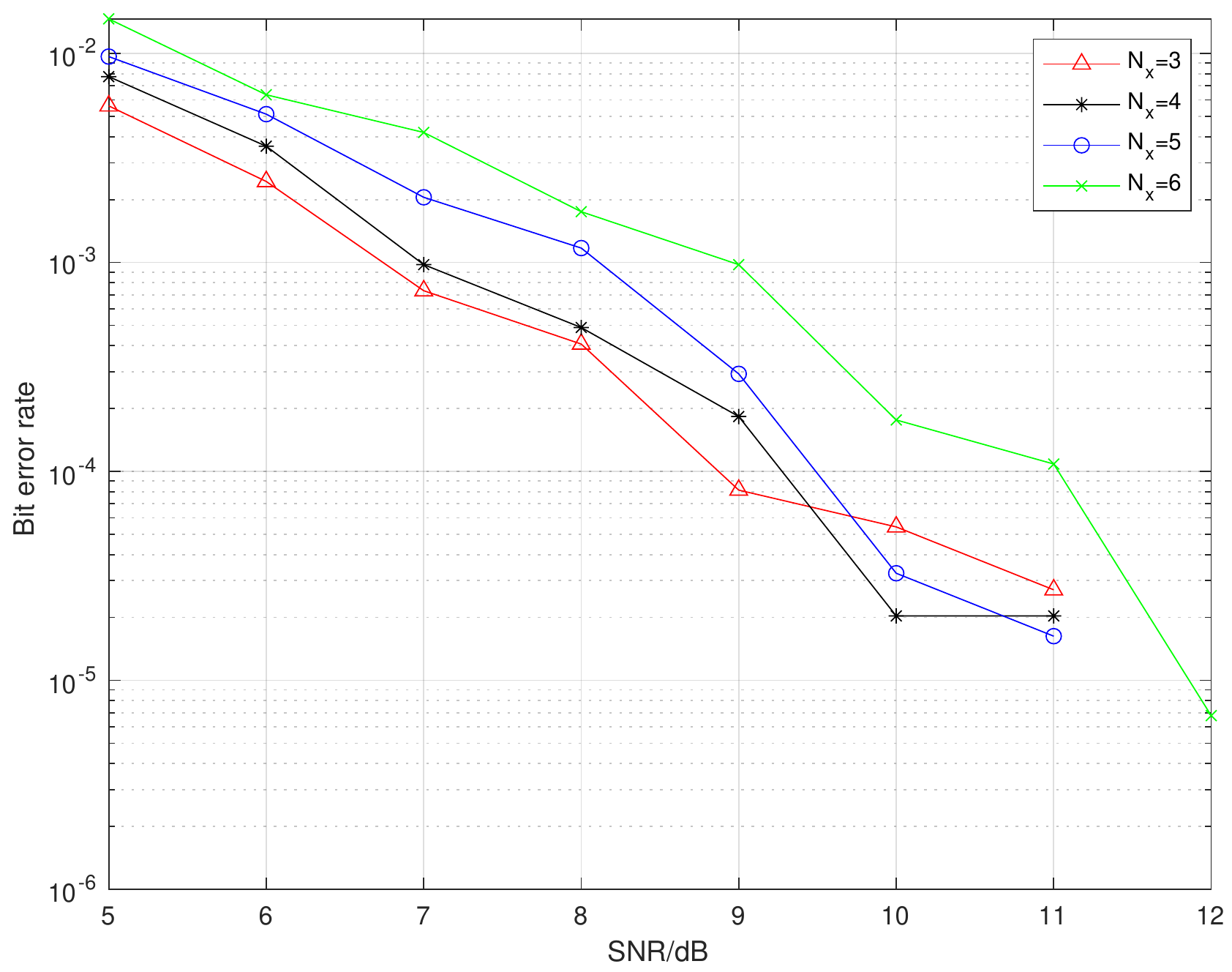}
\caption{BER versus SNR for data symbols obtained via the method of Sec.~\ref{alternative}.}
% on shared subcarriers with prior knowledge from private subcarriers on active antenna indices.}
\label{ber_pri}
\end{figure}

To evaluate the performance of the communication functionality we performed  Monte Carlo simulations and calculated the bit error rates from data symbols and antenna indices encoding  under different SNRs and  different numbers of active antennas.
Fig. \ref{ber} shows the performance of the communication system when applying the data symbol recovery method presented in Section~\ref{SSR}.
For a fixed SNR, the fewer the activated antennas the smaller the BER is. This is because  sparse signal recovery works better when the signal is sparser.

\begin{figure}%[!t]
\centering
\includegraphics[width=3in]{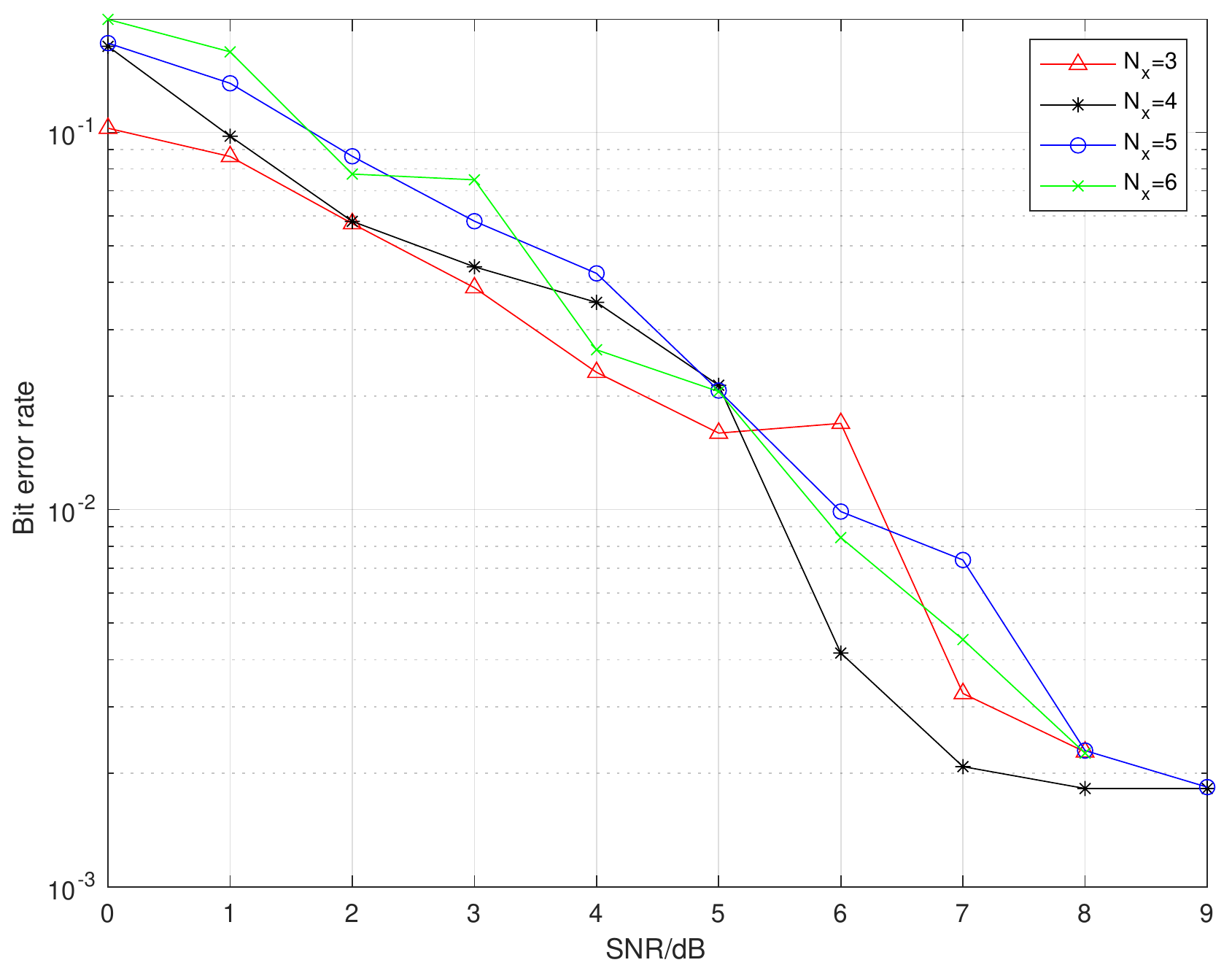}
\caption{BER versus SNR for GSM obtained via the method of Sec.~\ref{alternative}.}
\label{ber_pos}
\end{figure}

In Fig. \ref{ber_pri} we plot the BER  based on the received symbols,  when the communication receiver applies the symbol and active antenna indices recovery method described in Sec.~\ref{alternative}, i.e., 
the private subcarriers and active antennas indices are  identified first, and then the symbols are recovered via an  LS approach.
In comparison to Fig. \ref{ber}, one  can see that this approach achieves lower BER  for the same SNR and the same value of $N_x$. 
Indeed,  the use of private subcarriers not only enables the construction of a virtual array for the radar system, but it also makes the communication system more robust to noise.
The BER corresponding to the estimated antenna indices is shown in Fig.  \ref{ber_pos}.
  In the simulations, the position bit stream was randomly generated and mapped to a dictionary to decide the indices of active antennas.
  %
%  Fig.  \ref{ber_pos} shows the BER corresponding to antenna indices  encoding.
%  In the simulations, the position bit stream is randomly generated and mapped into a dictionary to decide the indices of active antennas. \textcolor{blue}{The active antenna indices were recovered based on the private subcarriers which are decided by the number of non-zero elements. After the sparse sensing, if the recovered vector contains only one non-zero element, the corresponding subcarrier will be labeled as private and the location of this non-zero element will reveal the index of active antenna which is assigned with this private subcarrier. Once all the private subcarriers are found, i.e., all active antennas are found, the position bits will be recovered by mapping the activation pattern back to the dictionary.} \textcolor{red}{EXPLAIN HOW - DO NOT JUST cite Sec.\ref{alternative} because here you d not use the symbols - if I understand correctly.}
One can see that the position encoding is robust to noise and the number of active antennas does not affect the result as in Fig. \ref{ber}. 

% The use of private subcarriers for active antenna index recovery enables a good performance of data symbol decoding on the shared subcarriers. \textcolor{red}{*****but you have not shown how easy it is to determine who is a private subcarrier}
%
% together promise a high bit rate with little errors.
% In the simulation, if the subarray of recovered indices is not the exactly the same as the chosen indices, it counts as failure.

Under the configuration provided in the table, the maximum bit rate of the system with no private subcarriers is $1.6398$ Gigabits per second, while the maximum bit rate of the same system with $N_x=5$ private subcarriers in every OFDM symbol is $1.6339$ Gigabits per second. 
% \textcolor{red}{did you use private subc in one symbo only? How many symbols did you use for Doppler?}\textcolor{blue}{XU: 256 OFDM symbols for Doppler} 
Thus, while the loss in bit rate from enabling private subcarrier is minor,  the improvement in BER is significant.
% The information from different position combination and activated subcarriers can have a comparable bit rate which enables the system to tradeoff the computation demand and the bit rate.  %\textcolor{violet}{SUN: MB/s stands for?? you need to give the definition}

\section{Conclusion}

We have proposed a novel MIMO-OFDM dual-function system using a sparse transmit array, whose active elements
are selected in a GSM fashion.
Most subcarriers are used in a shared fashion by the active antennas, except a set of subcarriers that are assigned to the transmit antennas in an exclusive fashion (private subcarriers).
 For the radar function, the system estimates angle, range and Doppler information using both private and shared subcarriers. The angle estimate is further improved by exploiting  a virtual array constructed based on the private subcarriers. The communication system can use the private subcarriers to estimate active antenna indices and thus decode spatial information. Subcarrier sharing allows for high communication rates. 
 The fact that only a small number of transmit antennas is active allows for low hardware cost of the DFRC system.

\end{document}